\documentclass[a4paper,11pt]{article}
\pdfoutput=1 

\usepackage{jinstpub} 

\title{\boldmath Study of the water Cherenkov detector with high dynamic range for LHAASO}

\author[a,b]{K. Jiang,}
\author[a,b,1]{Z. Tang,\note{Corresponding author.}}
\author[a,b,1]{X. Li,}
\author[a,b]{and C. Li}

\affiliation[a]{State Key Laboratory of Particle Detection and Electronics, University of Science and Technology of China, Hefei 230026, China}
\affiliation[b]{Department of Modern Physics, University of Science and Technology of China, Hefei 230026, China}

\emailAdd{zbtang@ustc.edu.cn}
\emailAdd{li124ste@ustc.edu.cn}

\abstract{The Large High-Altitude Air Shower Observatory (LHAASO) is being built at Haizi mountain, Sichuan, China at an altitude of 4410 m. One of its main goals is to survey the northern sky for very-high- energy (above 100 GeV) gamma ray sources via its ground-based Water Cherenkov Detector Array (WCDA). WCDA is 78000 $m^2$ in dimension and consists of 3120 water detector cells divided into 3 water ponds. A hemispherical 8-inch photomultiplier tube (PMT) CR365-02-1 from Beijing Hamamatsu Photon Techniques INC. (BHP) is installed at the bottom-center of each cell of the first water pond to collect Cherenkov light produced by air shower particles crossing water. This proceeding includes the technical design of WCDA, the design of a high dynamic range base for CR365-02-1, the PMT test system and test results of 997 PMTs.}

\keywords{Photon detectors for UV, visible and IR photons (vacuum); Large detector systems for particle and astroparticle physics; Cherenkov detectors}

\proceeding{International Conference on Instrumentaion for Colliding Beam Physics (INSTR-2020)\\
  February, 24-28,2020\\
  Budker Institute of Nuclear Physics, and Novosibirsk State University, Novosibirsk, Russia}

\begin{document}
\maketitle
\flushbottom

\section{The LHAAASO project and WCDA}
\label{sec:intro}
The Large High Altitude Air Shower Observatory (LHAASO)~\cite{Zhen2010LHAASO} is a unique and new generation cosmic ray station being built at the Haizi Mountain in Daocheng of Sichuan Province with an altitude up to 4400 meters above the sea level.
As shown in figure~\ref{fig:lhaaso}, the LHAASO experiment mainly includes three detector arrays~\cite{He2018Design}:
1 $km^2$ array (KM2A) composed of electromagnetic particle detectors (EDs) and muon detectors (MDs),
78,000 $m^2$ water Cherenkov detector array (WCDA), 
and 20 wide field-of-view air Cherenkov telescopes (WFCTA).
{
\begin{figure}[htbp]
\centering 
\includegraphics[keepaspectratio,width=0.7\textwidth]{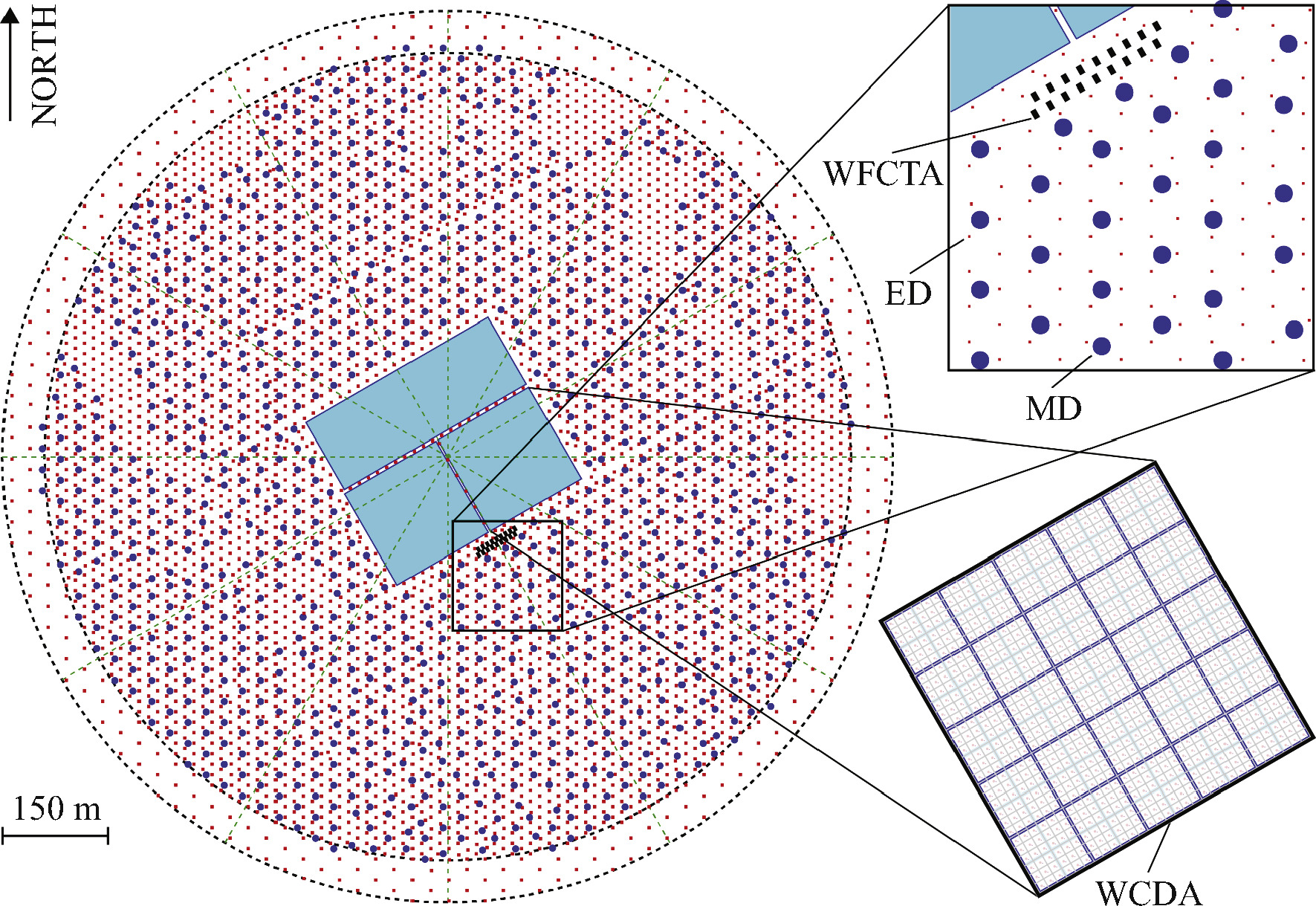}
\qquad
\caption{The layout of LHAASO detectors.} \label{fig:lhaaso}
\end{figure}
}

As one of the major detectors in the LHAASO project, the WCDA focuses on surveying the northern sky for gamma ray sources at the energy between 100 GeV and 20 TeV. 
Figure~\ref{fig:wcda} shows the schematic diagram of the overall layout of WCDA.
It consists of 3 water ponds and covers an effective area of 78000 $m^2$ with an effective water depth of 4.5 m.
Each water pond is divided into 5 m $\times$ 5 m detector cells partitioned by black curtains to prevent penetration of the light from neighboring cells.
The first two water ponds with an effective area of 150 $\times$ 150 $m^2$ contain 900 detector cells each. 
The third water pond with an area of 300 $\times$ 110 $m^2$  contains 1320 detector cells.
There are 3120 detection cells in total. 
A hemispherical 8-inch photomultiplier tube (PMT) CR365-02-1 from Beijing Hamamatsu Photon Techniques INC. (BHP) is installed at the bottom-center of each cell of the first water pond to collect Cherenkov light produced by air shower particles crossing water.
A 1.5-inch PMT is placed aside each 8-inch PMT to extend the dynamic range. 
{
\begin{figure}[htbp]
\centering 
\includegraphics[keepaspectratio,width=0.8\textwidth]{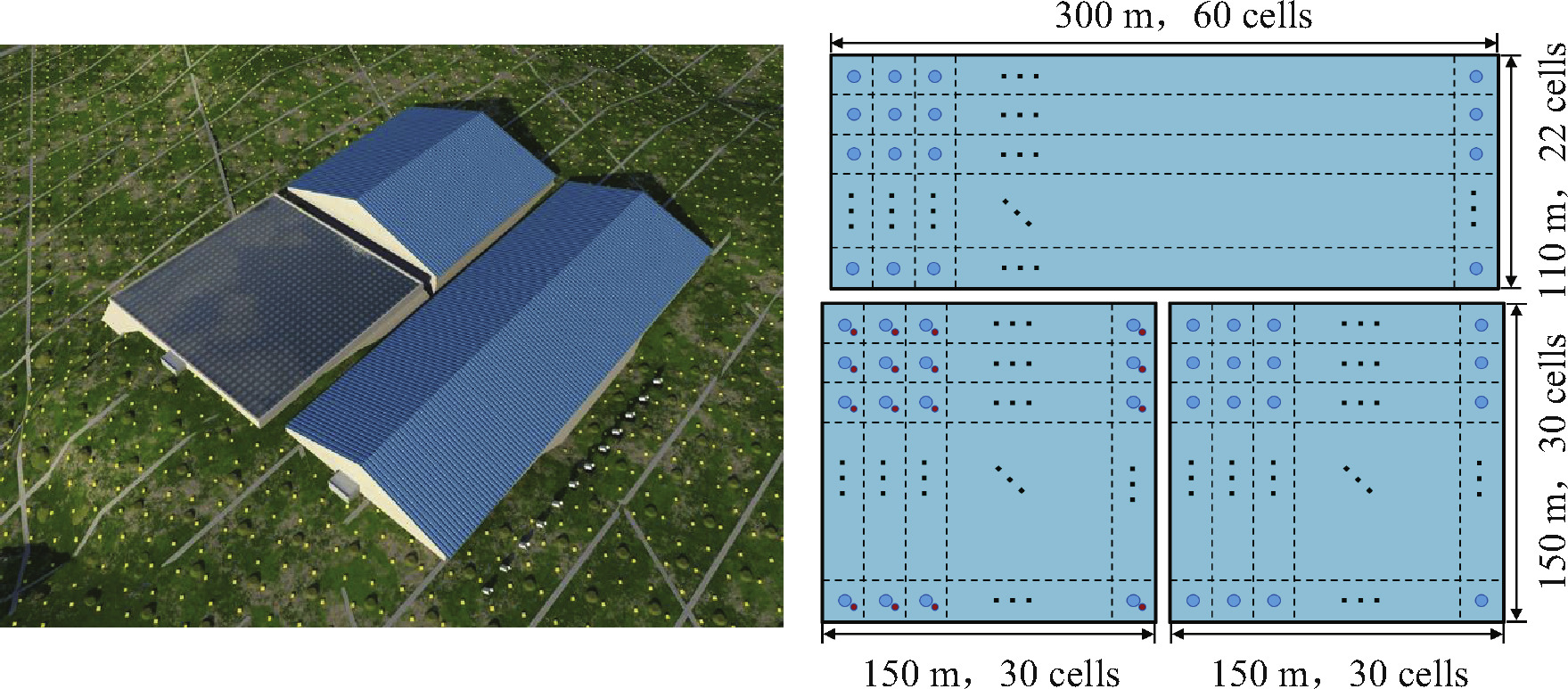}
\qquad
\caption{The schematic diagram of the overall layout of WCDA.} \label{fig:wcda}
\end{figure}
}

\section{PMT CR365-02-1 and base circuit design}
\label{sec:pmtandbase}
{
\begin{table*}
\centering
  \begin{tabular}{c|c}
    Parameters  &  Specifications  \\\hline
    Distribution of working voltage & Mean $\pm$  100 V, $<$ 2000 V \\ 
    Distribution of HV-gain coefficient ($\alpha$) &   Mean $\pm$ 0.5  \\
    Peak-to-valley ratio   &  $>$ 2.0  \\
    Quantum efficiency  &  $>$ 22\%  \\
    Transit time spread (FWHM)  &  $<$ 4 ns  \\
    Dark count rate ($>1/3$ PE)   &   $<$ 5 kHz \\
    After-pulse rate ($100-10000~\textrm{ns}$)   &   $<$ 5\% \\
    Anode linearity (5\%)    &   $>$ 1000 PEs\\
    Dynode linearity (5\%)   &   $>$ 4000 PEs\\
    Distribution of anode-to-dynode charge ratio  &Mean $\pm$ 15\%\\
\end{tabular}
	\caption{Specification requirements of CR365-02-1. The working gain is $3\times10^6$.}
  \label{tab:cut}
\end{table*}
}
According to the Monte Carlo simulations on gamma ray from Crab nebula, large-sized PMTs for WCDA are required to have good single photoelectron (SPE) resolution, large linearity up to 4000 photoelectrons (PEs) and fast timing characteristics~\cite{wcdaicrc2011,Performancewcdaprototype}. 
The specifications for 8-inch PMTs are listed in Table~\ref{tab:cut}.
A special high dynamic range base circuit for PMT CR365-02-1 is designed to meet these requirements(see figure~\ref{fig:base}).
Signals from PMT are read out from two outputs: one from the anode and the other one from the 8-th dynode~\cite{Huang2013R5912,Zhao2016XP1805}. 
The ratio of the gain of the two outputs is specially tuned to about 50 to balance the dynamic range and overlapping range. 
The voltage distribution on the electrodes of the PMT is tapered in order to improve the linearity of the PMT. 
The damping resistor connected to the 8-th dynode has been fine tuned to minimize the oscillation of the pulse. 
The assembly of the CR365-02-1 is shown in figure.~\ref{fig:base}. 
There are 3 30-meters-long cables, one for high voltage supply and two for signal outputs, in bundle covered by waterproof material. 
{
\begin{figure}[htbp]
\begin{center}
\includegraphics[keepaspectratio,width=0.37\textwidth]{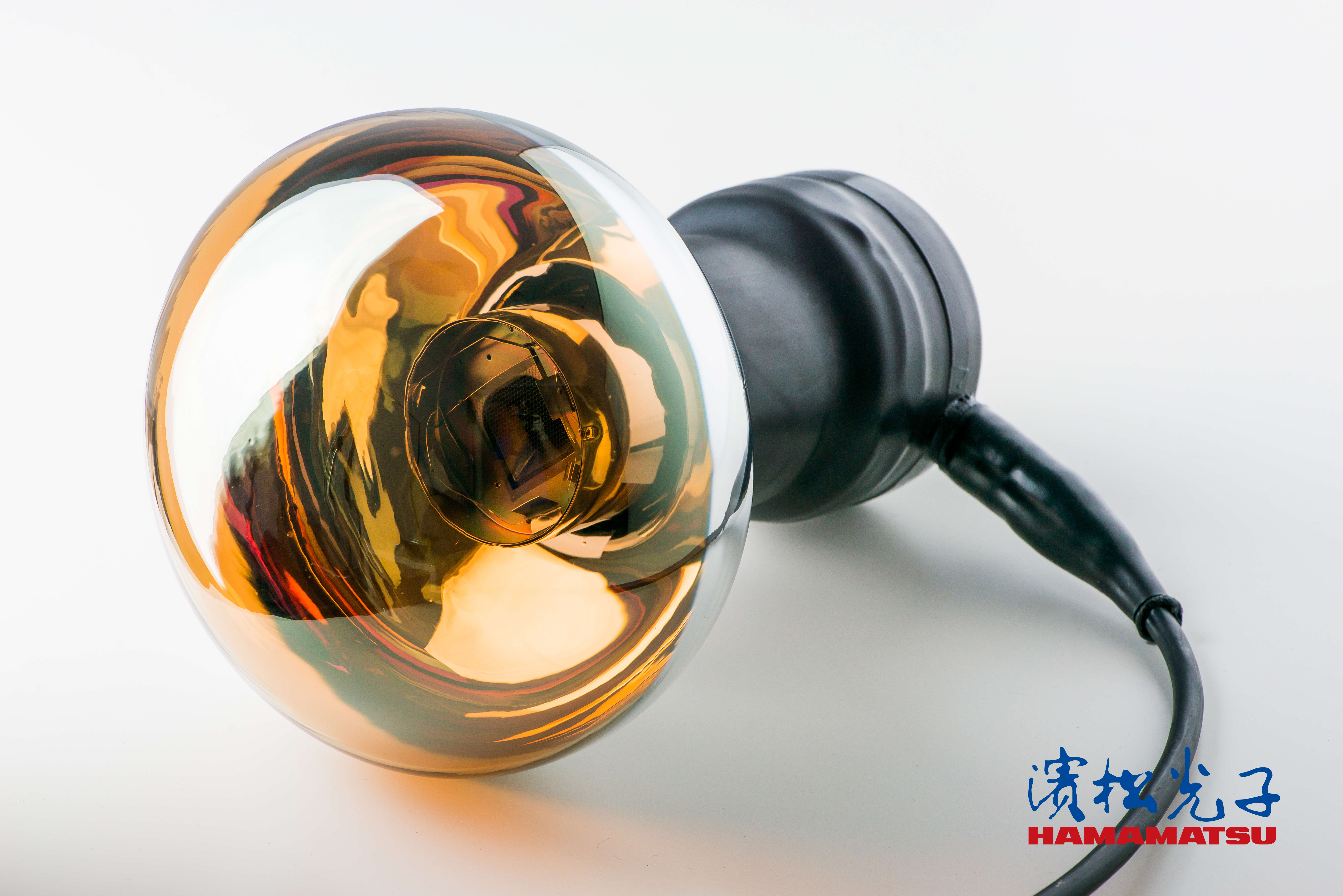}
\includegraphics[keepaspectratio,width=0.5\textwidth]{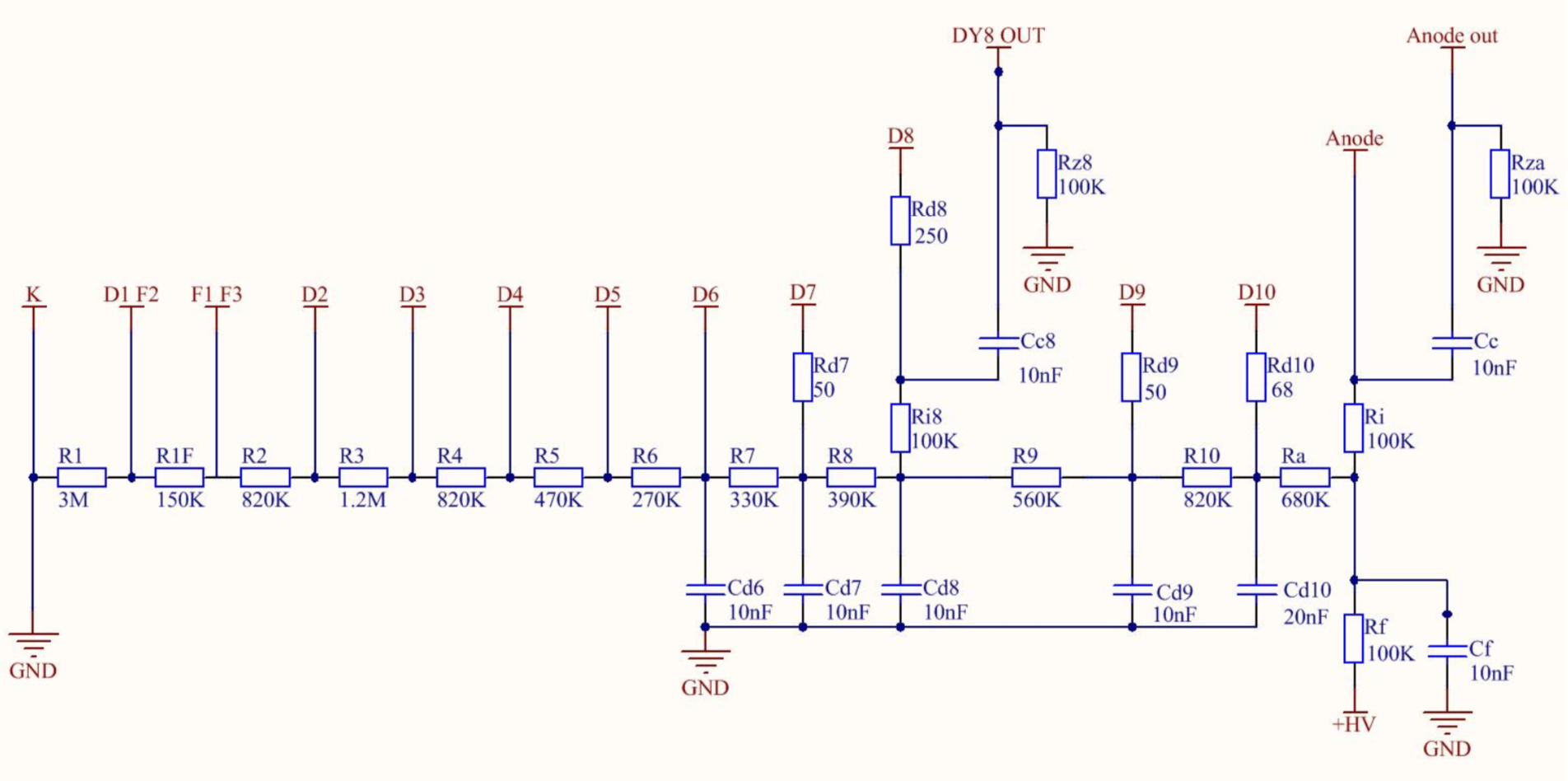}
\vspace*{+0mm}
\caption{The PMT CR365-02-1 from BHP and the base circuit used for the LHAASO WCDA.} \label{fig:base}
\end{center}
\end{figure}
}

\section{The PMT test system and test results}
\label{sec:pmttestsystem}
{
\begin{figure}[htbp]
\begin{center}
\includegraphics[keepaspectratio,width=0.66\textwidth]{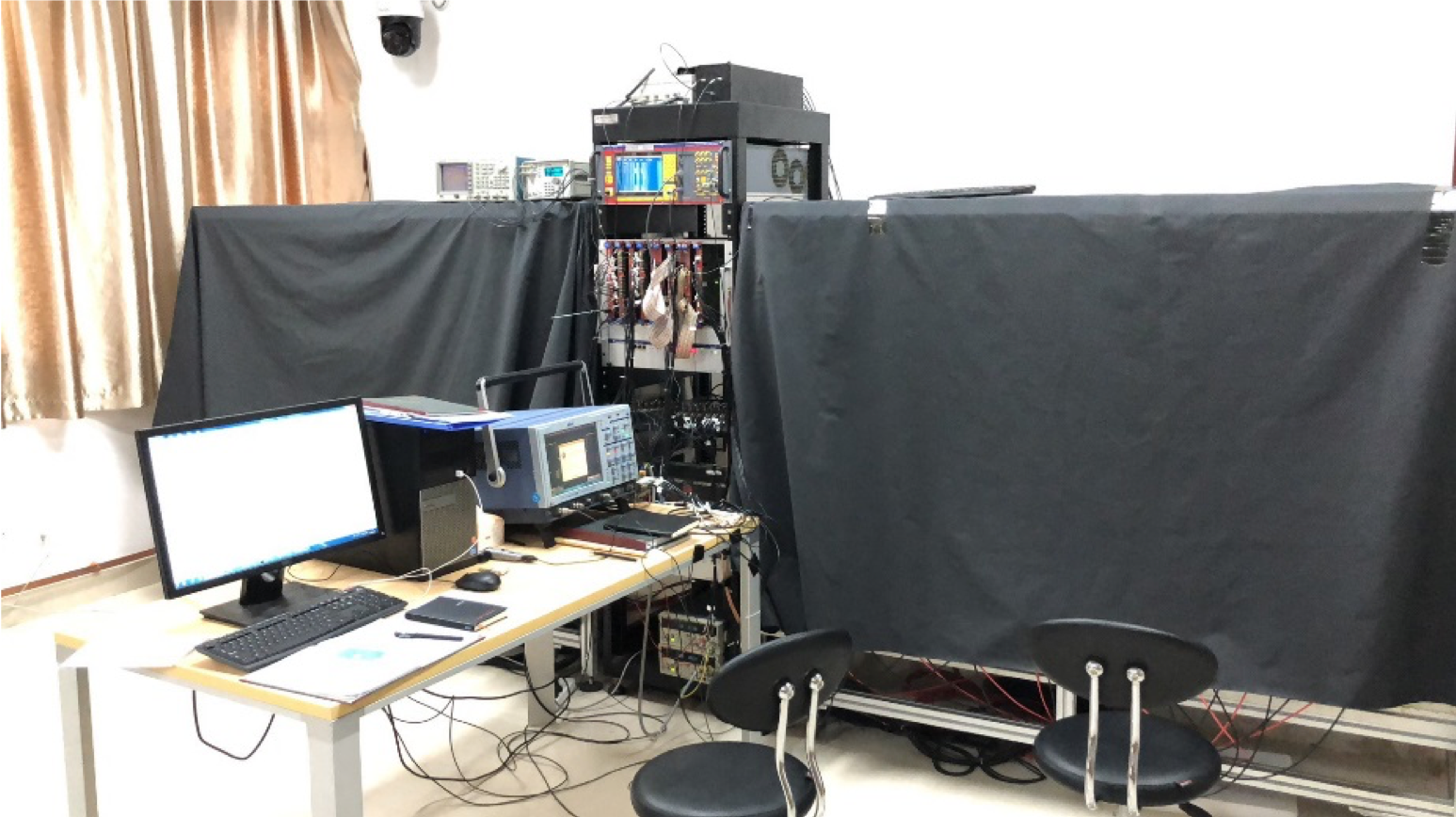}
\vspace*{+0mm}
\caption{The PMT test facility.} \label{fig:facility}
\end{center}
\end{figure}
}
The PMT test system is setup at University of Science and Technology of China in Hefei.
It consists of dark boxes, light sources, front end electronics , and data acquisition (DAQ) system.
The system is designed to test 16 PMTs simultaneously during one test run.
Two calibrated PMTs stay in the setup to server as reference and also monitor the stability of the whole system. 
The measured PMT parameters include: high voltage response, SPE spectrum, transit time spread (TTS), relative quantum efficiency, dark noise rate, after pulse rate~\cite{Zhao2016Afterpulse}, nonlinearity and anode to dynode charge ratio.  
The test procedure is pretty much automatic. 
Each test run takes about 15 hours and the system can test 14 new PMTs per day. 
Until recently, 997 PMTs have been tested.

{
\begin{figure}[htbp]
\begin{center}
\includegraphics[keepaspectratio,width=0.49\textwidth]{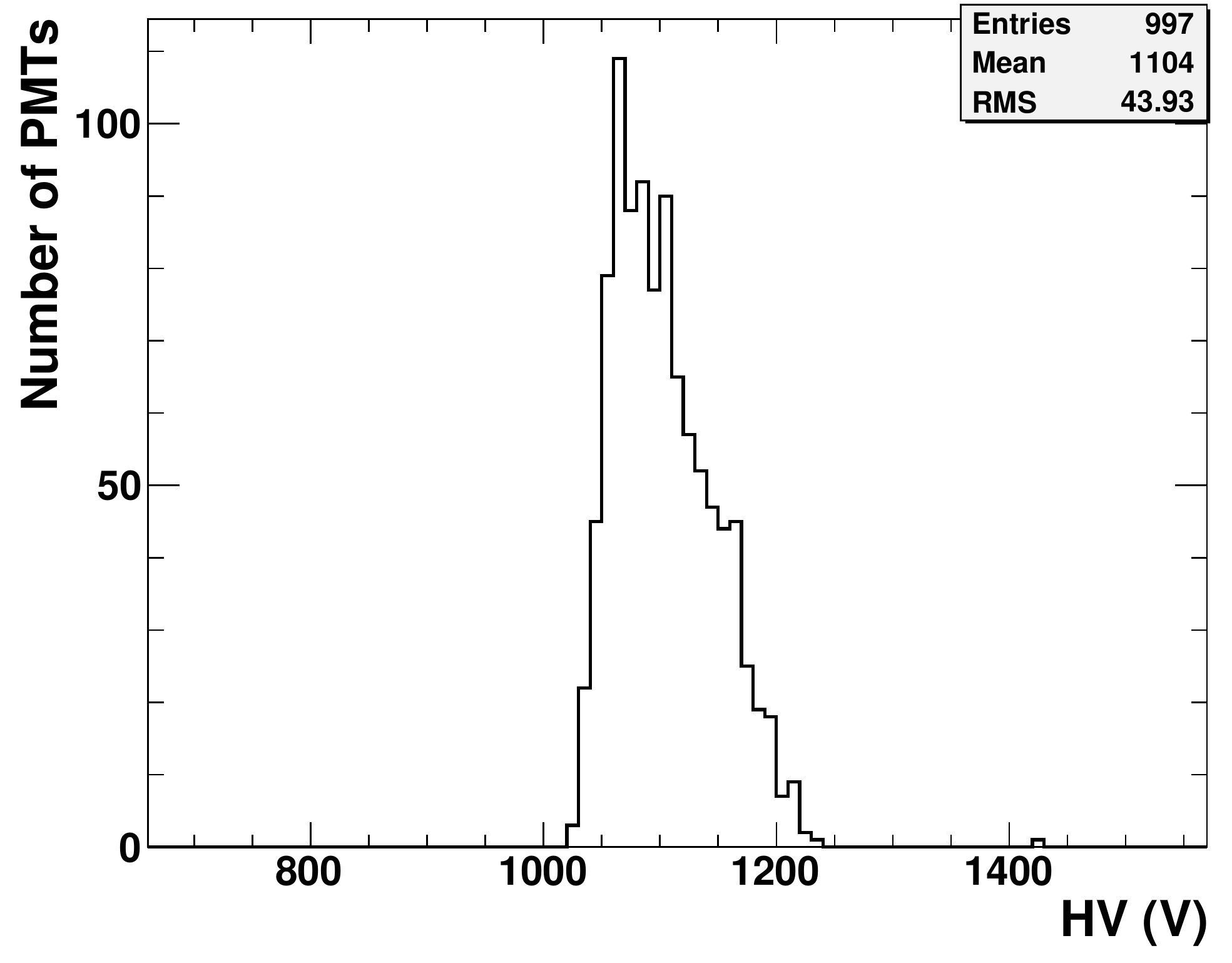}
\vspace*{+0mm}
\caption{Distribution of the working high voltage $HV_{opt}$ for a gain of $3\times10^6$. } \label{fig:hv}
\end{center}
\end{figure}
}
{
\begin{figure}[htbp]
\begin{center}
\includegraphics[keepaspectratio,width=0.3\textwidth]{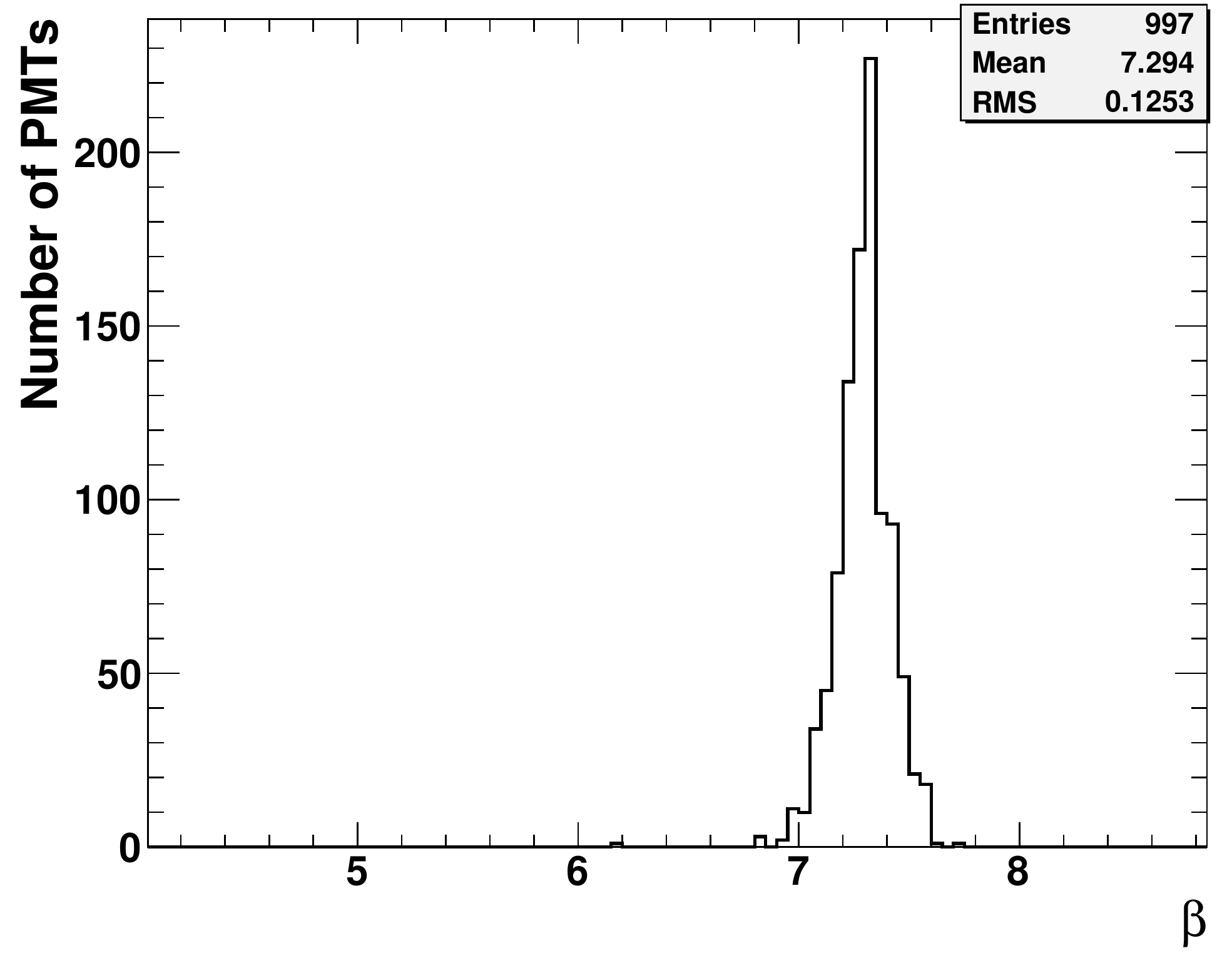}
\includegraphics[keepaspectratio,width=0.3\textwidth]{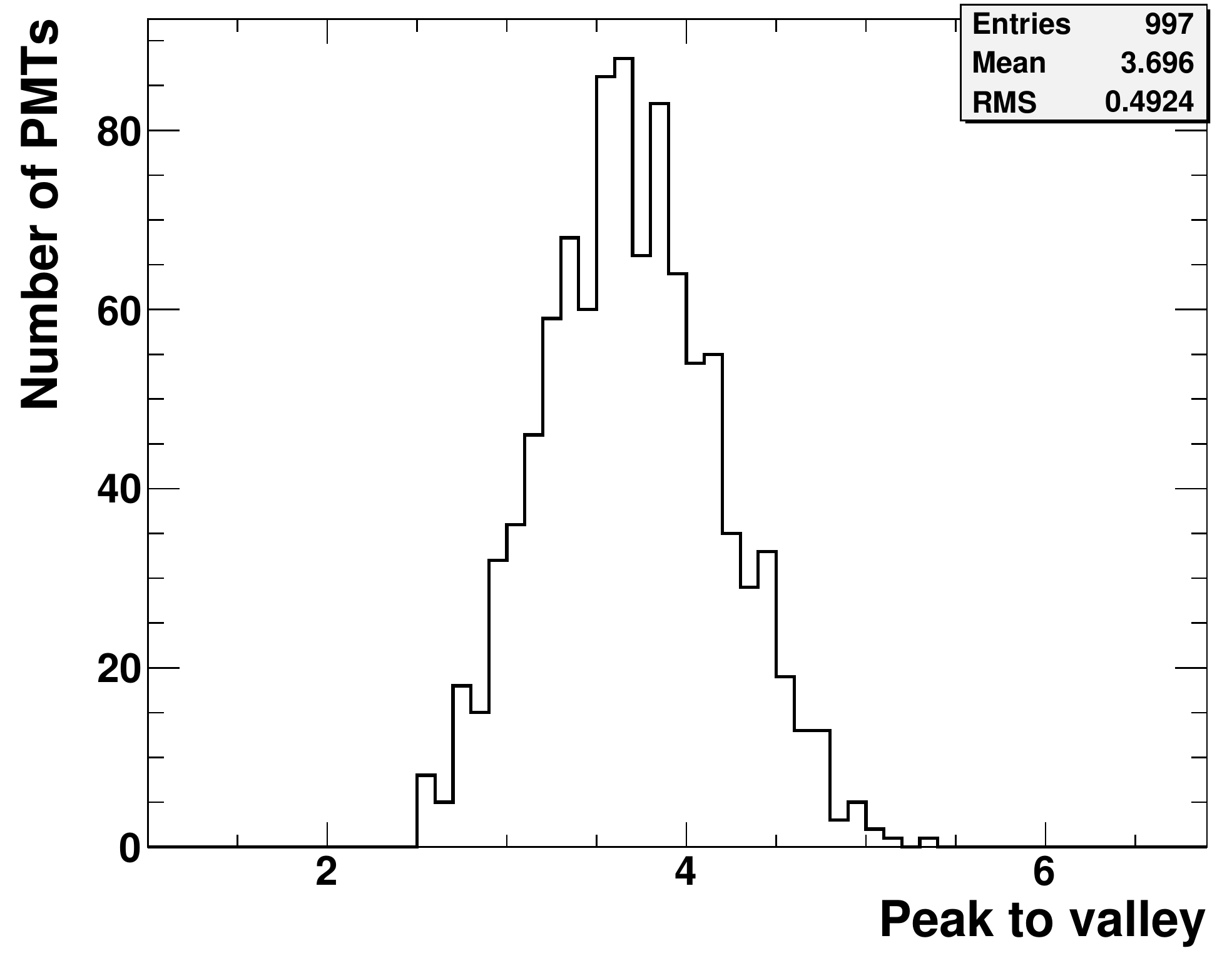}
\includegraphics[keepaspectratio,width=0.3\textwidth]{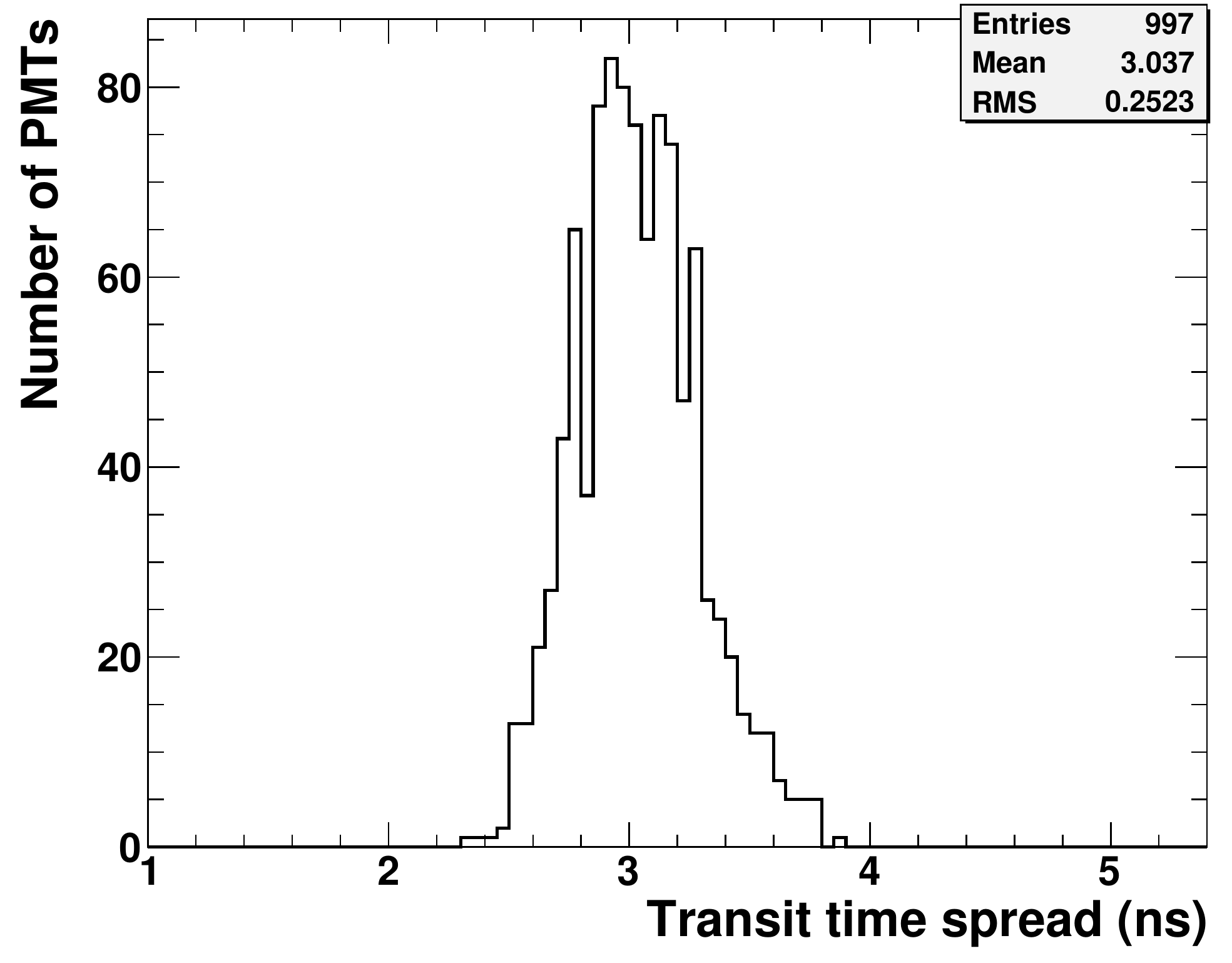}
\includegraphics[keepaspectratio,width=0.3\textwidth]{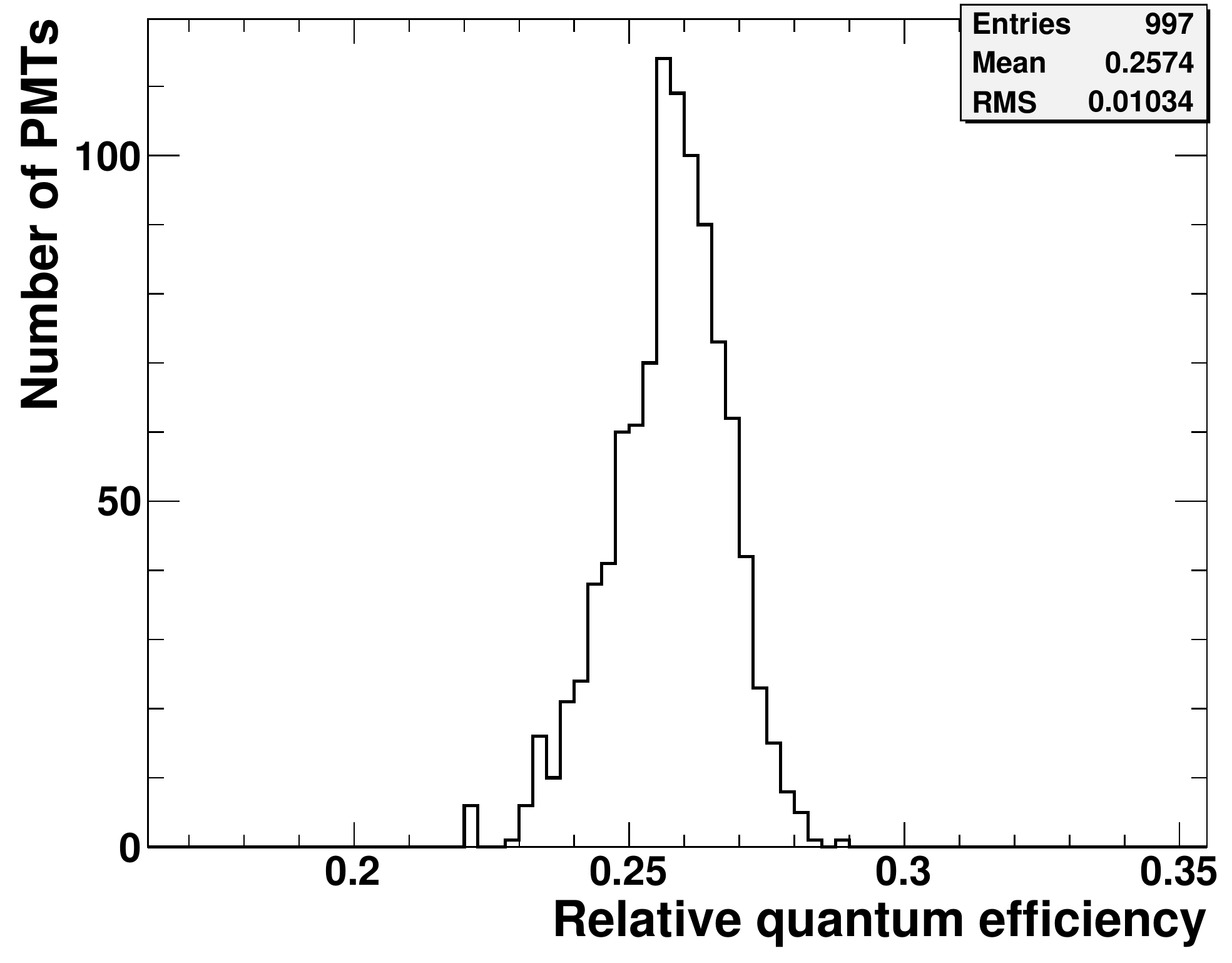}
\includegraphics[keepaspectratio,width=0.3\textwidth]{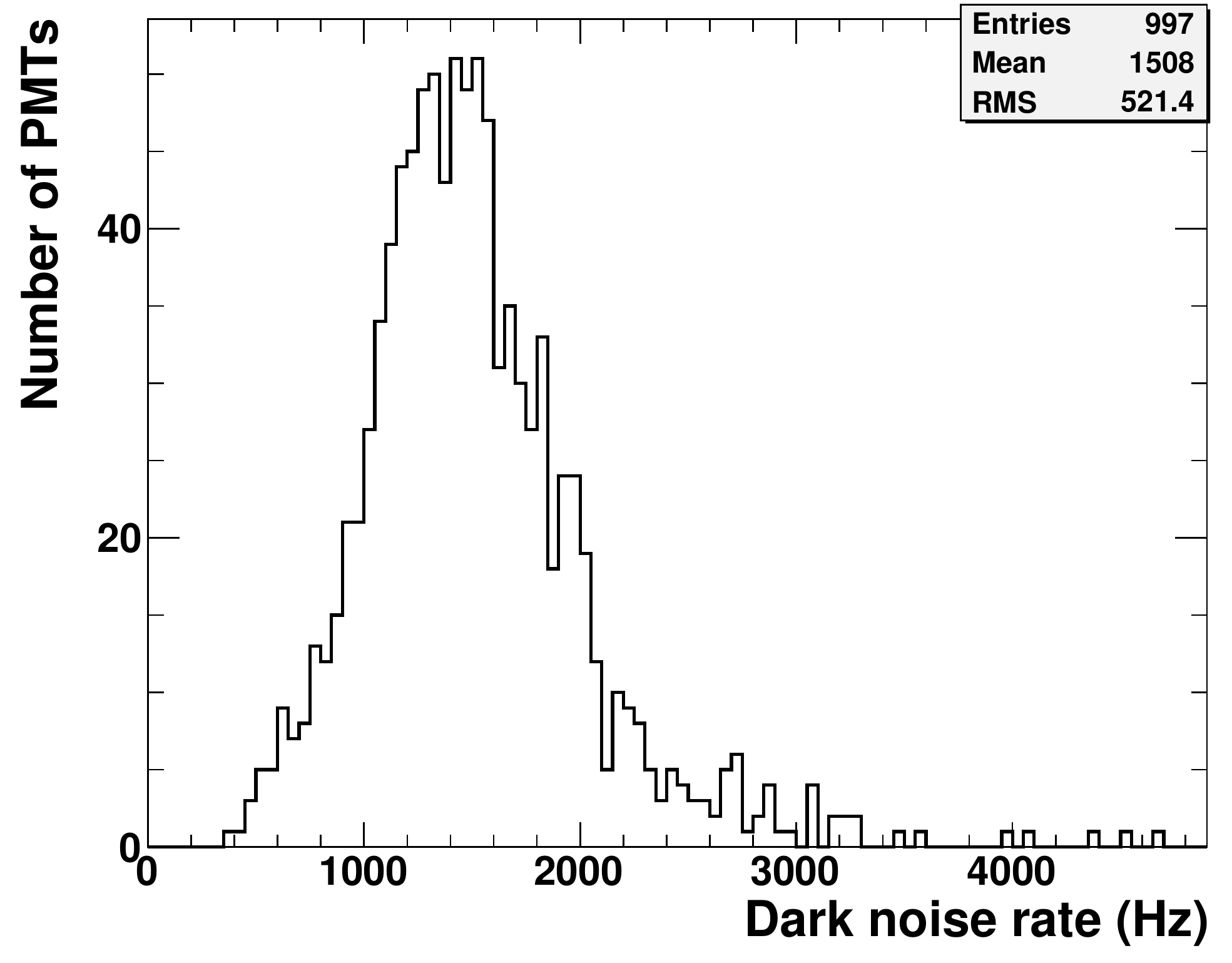}
\includegraphics[keepaspectratio,width=0.3\textwidth]{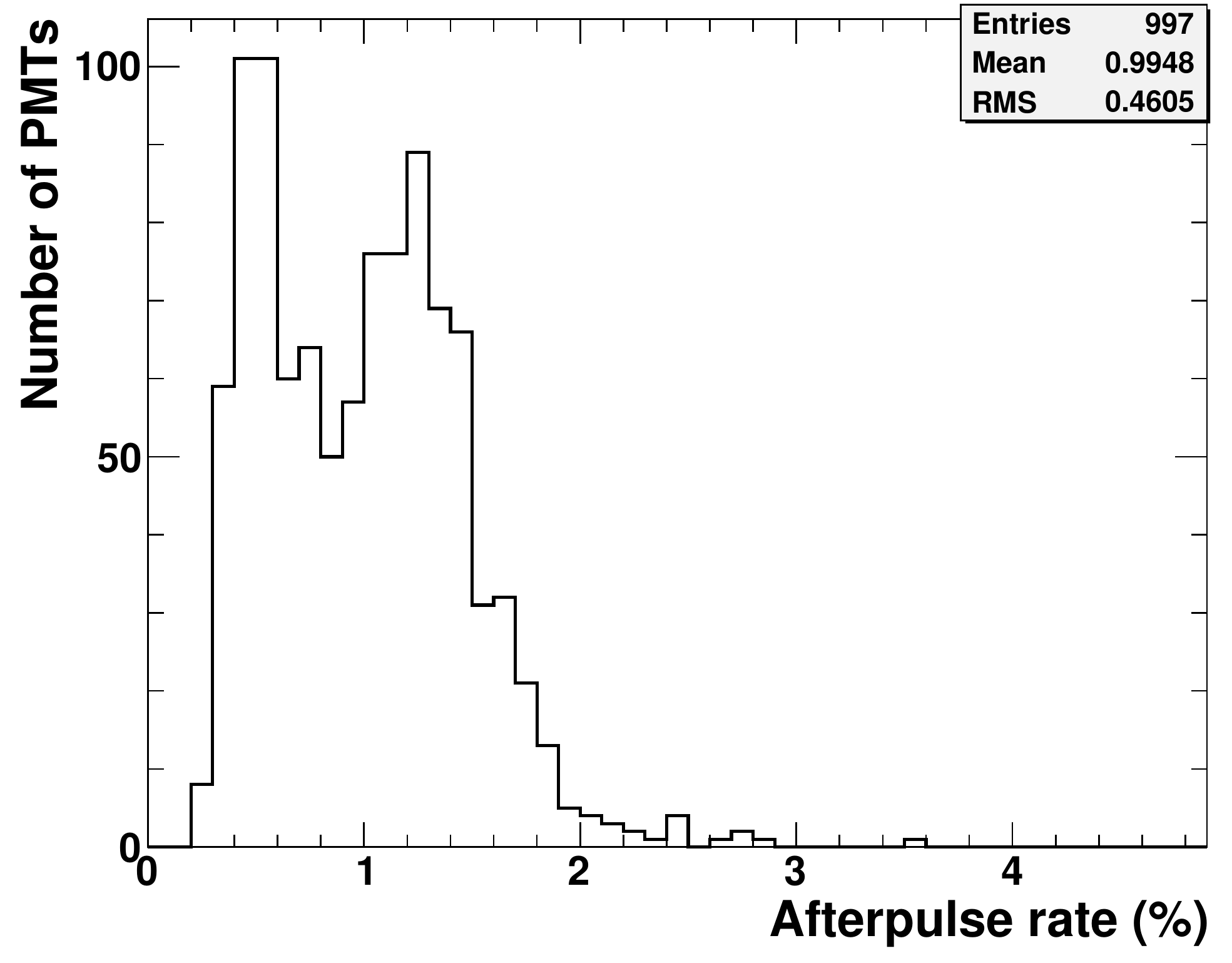}
\includegraphics[keepaspectratio,width=0.3\textwidth]{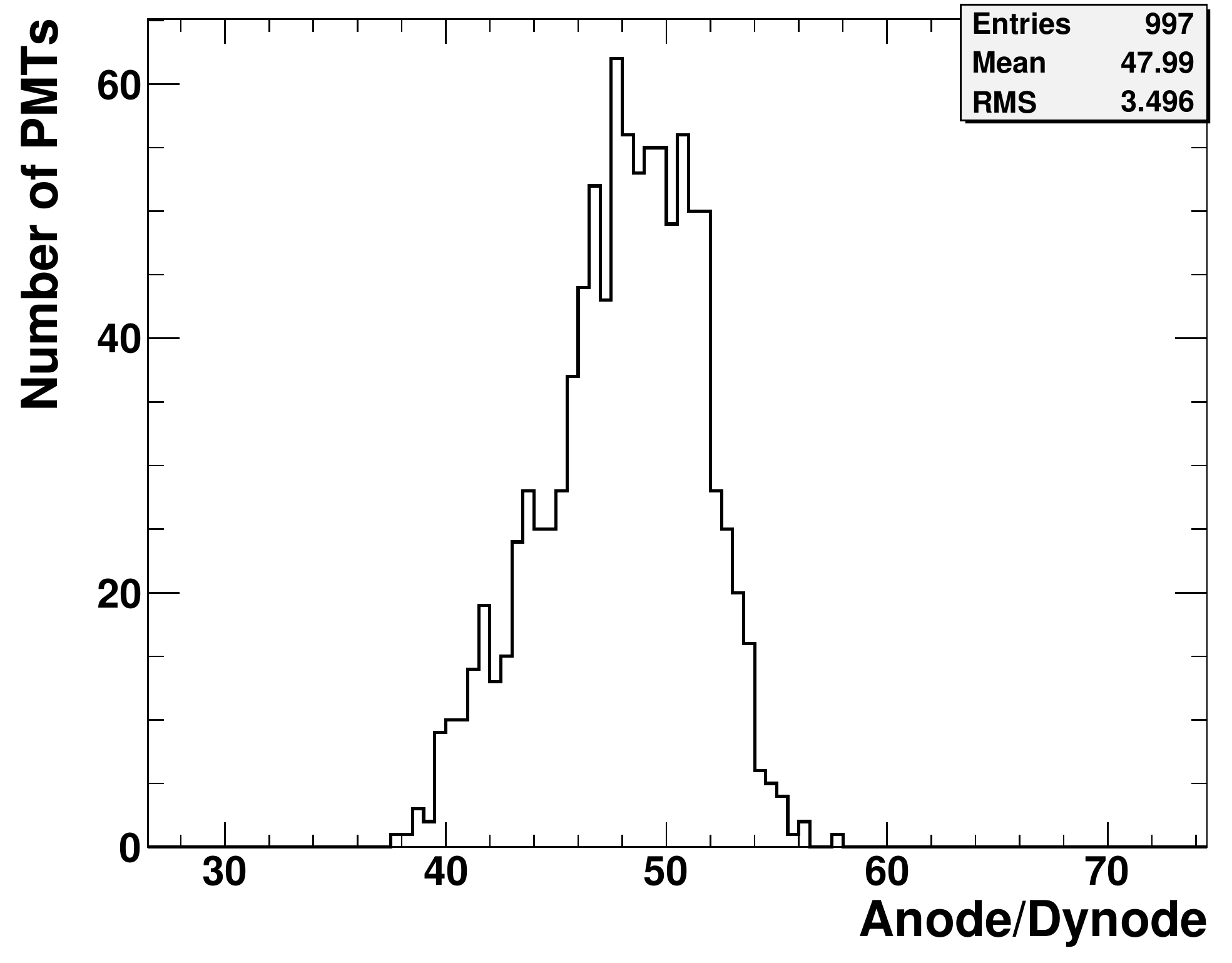}
\includegraphics[keepaspectratio,width=0.3\textwidth]{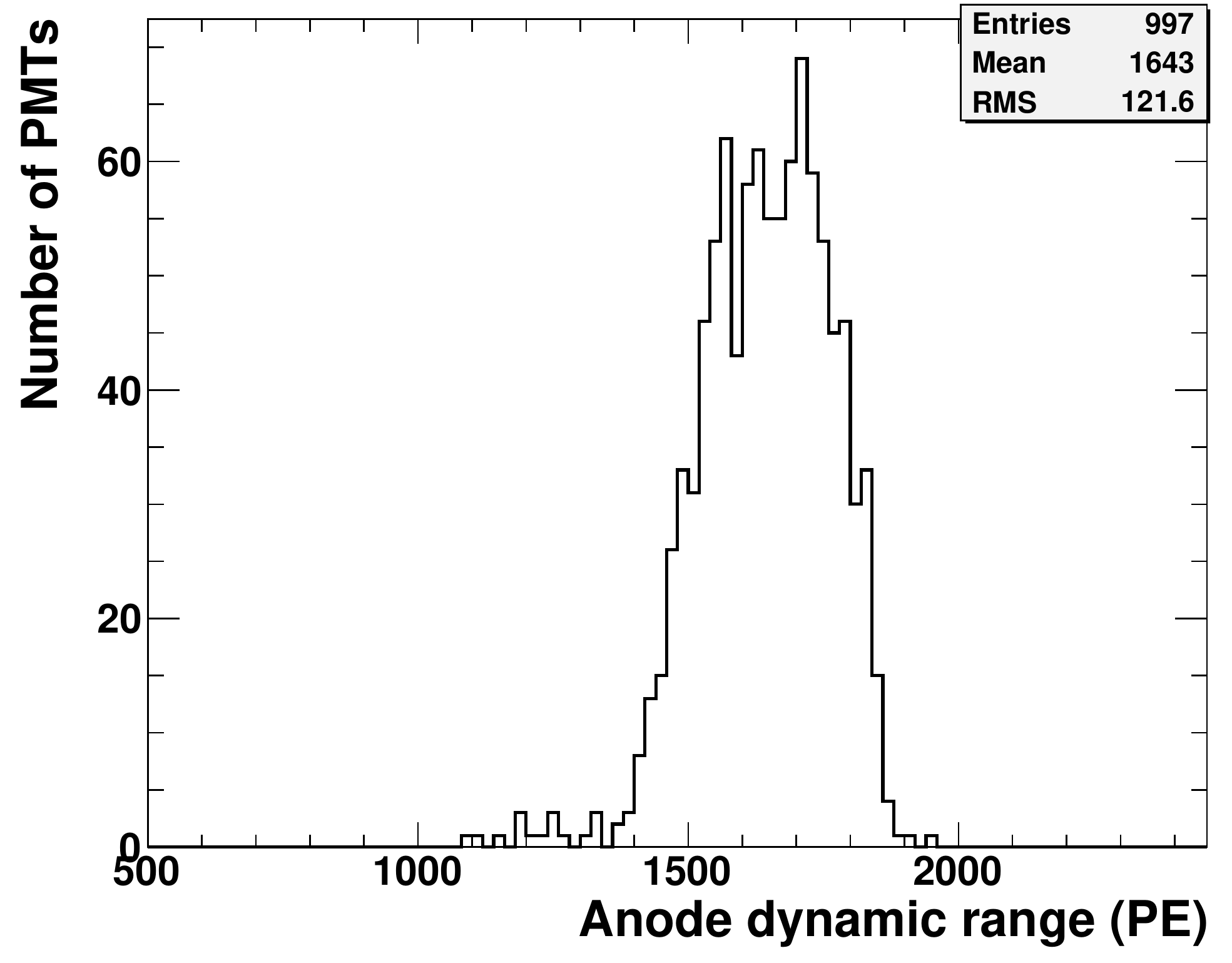}
\includegraphics[keepaspectratio,width=0.3\textwidth]{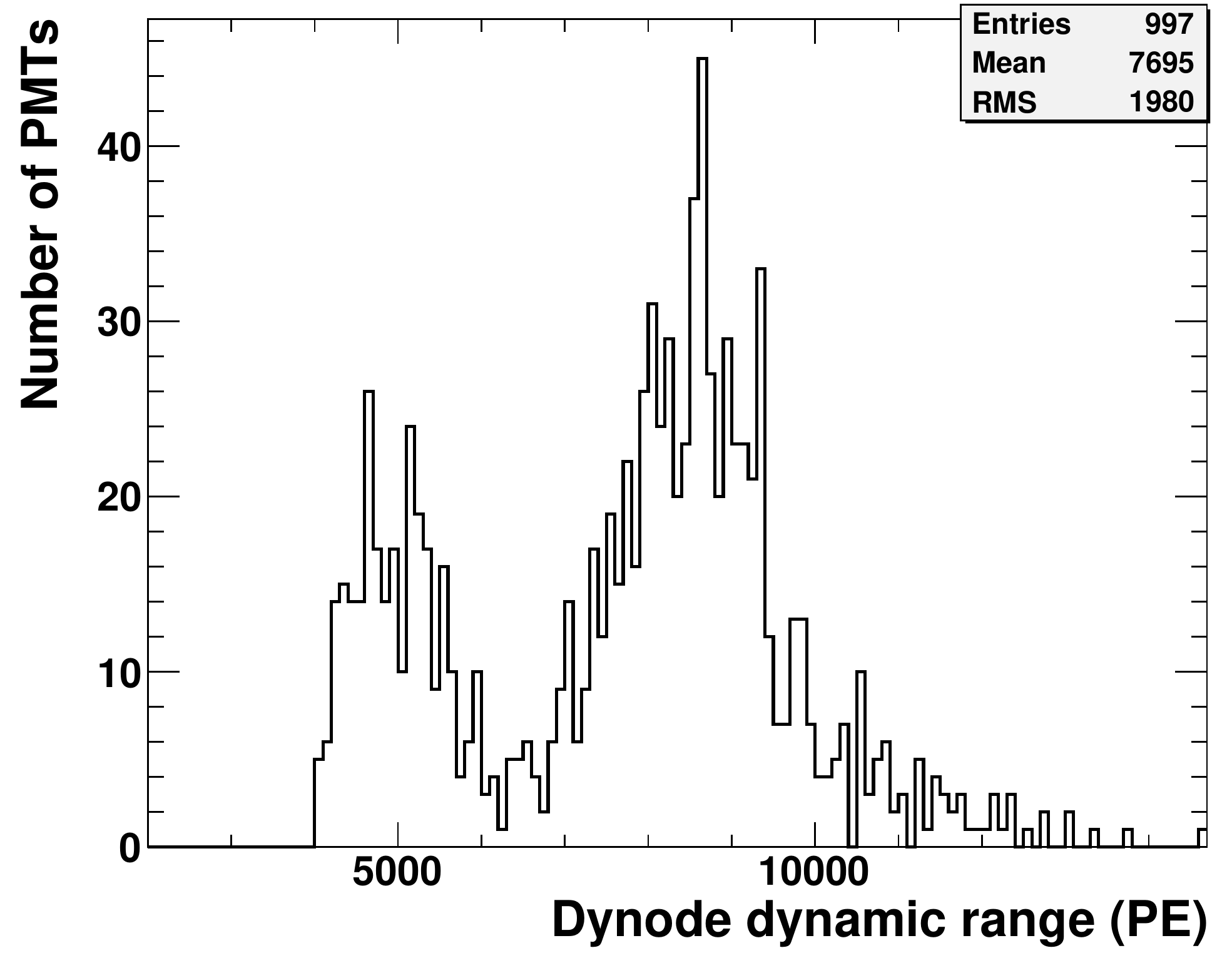}
\vspace*{+0mm}
\caption{Distributions of the measured PMT parameters at a gain of $3\times10^6$. } \label{fig:otherpara}
\end{center}
\end{figure}
}
Figure.~\ref{fig:hv} shows the distribution of the working high voltage $HV_{opt}$ for a gain of $3\times10^6$. 
The mean value of $HV_{opt}$ is 1104 V. 
17 PMTs (1.7\%) failed to conform the specification for $HV_{opt}$ ($mean\pm100$ V).
Due to a relative long tail on the right hand side of the $HV_{opt}$ distribution,
all failed PMTs have a $HV_{opt}$ larger than 1204 V.
The distributions of the other measured parameters at a gain of $3\times10^6$ are shown in figure.~\ref{fig:otherpara}. 
Among all the tested PMTs, one PMT failed in conforming the specification for amplification voltage coefficient $\beta$ ($mean\pm0.5$),
17 PMTs (1.7\%) failed in $HV_{opt}$ ($mean\pm100$ V),
32 PMTs (3.2\%) failed in A/D ($mean\pm15\%$).
4 PMTs failed in both $HV_{opt}$ and A/D.

\section{Summary}
\label{sec:summary}
To meet the requirements of LHAASO-WCDA, a PMT voltage divider circuit with high dynamic range has been designed.
A PMT batch test system for LHAASO project is designed and built.
Until now, a total of 997 8-inch CR365-02-1 have been  tested for LHAASO-WCDA, with 46 (4.6\%) unqualified. 


%


\acknowledgments
The research presented in this proceeding has received strong support from LHAASO collaboration and National Natural Science Foundation of China (No. 11675172, 11775217).
We would like to specially thank Prof. Zhen Cao, Huihai He, Zhiguo Yao, Mingjun Chen, Bo Gao and other members of the LHAASO Collaboration for their valuable support and suggestions.


\bibliographystyle{unsrt}
\bibliography{jinst-latex-sample}

%
%
%


\end{document}